\documentclass[%
11pt, tightenlines,
onecolumn,
superscriptaddress,
 amsmath,amssymb,
aps,
prl,
floatfix,
letterpaper,
notitlepage
]{revtex4-1}

\usepackage{amsmath,amssymb,amscd,amsxtra}

\usepackage{graphicx}
\usepackage{epstopdf}

\usepackage{color}

\usepackage{subfig}

\newcommand\Rey{\mbox{\textit{Re}}}

\newcommand\bu{\mathbf{u}}
\newcommand\ubar{\bar{\mathbf{u}}}

\newcommand\etal{\mbox{\textit{et al. }}}

\begin{document}

\title{Marine crustaceans with hairy appendages: role of hydrodynamic boundary layers in sensing and feeding}

\author{Kaitlyn Hood}
 \email{kaitlyn.t.hood@gmail.com}
\affiliation{%
 Department of Mechanical Engineering,
 Massachusetts Institute of Technology,
 Cambridge, MA 02139, USA
}%

\author{M. S. Suryateja Jammalamadaka}
\affiliation{%
 Center for Research and Interdisciplinarity, 75005 Paris, France
}%
\affiliation{%
 Department of Mechanical Engineering, Massachusetts Institute of Technology,
 Cambridge, MA 02139, USA
}%

\author{A. E. Hosoi}
\affiliation{%
 Department of Mechanical Engineering,
 Massachusetts Institute of Technology,
 Cambridge, MA 02139, USA
}%

\date{\today}

\begin{abstract}
      Decapod crustaceans have appendages with an array of rigid hairs covered in chemo-receptors, used to sense and track food. Crustaceans directly influence the flow behavior by changing the speed of flow past the hairy surface, thereby manipulating the Reynolds number ($\Rey$). Hairs act either as a rake – diverting flow around the hair array, or as a sieve – filtering flow through the hairs. In our experiments, we uncover a third transitional phase: deflection – where the flow partially penetrates the hair array and is deflected laterally. We develop a reduced order model that predicts the flow phase based on the depth of the boundary layer on a single hair. This model with no fitting paramaters agrees very well with our experimental data. Additionally, our model agrees well with measurements of both chemo-sensing and suspension-feeding crustaceans, and can be generalized for many different geometries.
\end{abstract}

\maketitle


Hairy surfaces submerged in fluid flow are ubiquitous in nature, from crustacean olfaction \cite{koehl2001lobster,koehl2001fluid}, suspension feeding \cite{conova1999role}, to nectar drinking \cite{nasto2018viscous}. While we can define a hairy surface in simple terms -- an array of filaments anchored at one end to a surface -- the utility and observed behavior of hairy surfaces is as varied as the organisms that employ them.

For example, the olfactory organ of decapod crustaceans consists of an array of sensilla, called aesthetascs, on the lateral flagellum of the antennule. Each aesthetasc contains olfactory receptor neurons. This system can be viewed as an array of rigid hairs attached to an antenna, which is then flicked through the water in order to sample odors.

The motion of marine organisms at micrometric scales is different from human scales. At small scales, the viscous drag on an organism vastly outweighs inertia. This phenomenon is indicated by the Reynolds number, $\Rey$, a dimensionless ratio of inertial to viscous stresses. However, flow past hairy surfaces is not limited to the low-$\Rey$ regime. Crustacean olfaction and suspension feeding both occur at $\Rey = O(1)$. Koehl \etal  observed that crustaceans flick their antennae at different speeds, intentionally manipulating the Reynolds number $\Rey$ to achieve different states of flow. The two states observed are called \textit{rake} -- where the fluid inside the bed of hairs is stagnant, and \textit{sieve} -- where the fluid inside the bed of hairs travels opposite the motion of the antenna. Koehl \etal have conjectured that crustaceans use these different phases of flow to aid olfaction. On the slow down-stroke, the hair bed collects a stagnant packet of fluid, allowing chemicals to diffuse to the chemo-receptors on the hairs. Conversely, on the fast up-stroke, the hair bed releases the packet of fluid in order to sample a new packet. 

Recent technological advances in fabrication have allowed engineers to implement hairy surfaces in devices. For example, a bed of flexible hairs anchored at an angle to the surface impedes flow in one direction \cite{alvarado2017nonlinear}, acting as a steady-state hydrodynamical diode. Robotic pollinators use hairs coated in a gel to transport pollen \cite{chechetka2017materially,amador2017sticky}. 
Furthermore, passive and/or steady-state devices are highly valued for their simplicity and robustness. Hairy surfaces, then, are an ideal design element to study, for their ubiquity, range of exhibited behavior, and robustness. 

Increasingly, inertia is used as a design element in microfluidic and hydrodynamic devices -- for particle filtering \cite{DiCarlo09,DiCarlo11}, flow cytometry \cite{DiCarlo10}, and entrapment of live cells in droplets for tissue printing \cite{edd2008controlled,amini2017inertial}. 
While the equations of motion for fluids with $\Rey = O(1)$ are nonlinear and therefore difficult and costly to solve, there has been an emergence of asymptotic and numerical theories to enable rational design with inertia \cite{klotsa2009chain, hood15inertial, true2017hydrodynamics}. However, in this regime, results are  sensitive to variations in the boundary conditions, and so far theories have been ad-hoc. Inertial flow over complicated and intricate surfaces is an open field for study.

To close this gap between biology and rational design, we investigate a bio-inspired model system of rigid hairs subject to inertial flow at $\Rey = O(1)$. In order to use inertia as a design element, similar to the olfaction mechanism in crustaceans, we need a theory for the flow phase based on experimental parameters such as hair length, diameter, and spacing length. Intuitively, the flow phase should be determined by the depth of the boundary layer on the hairs: rakes arise from large overlapping boundary layers and sieves arise from small boundary layers. We develop a quantitative theory for predicting the depth of the boundary layer, which can then be used to design a hair bed with desired flow phase.

\begin{figure}[t]
	\centering
	\includegraphics[scale = 1] {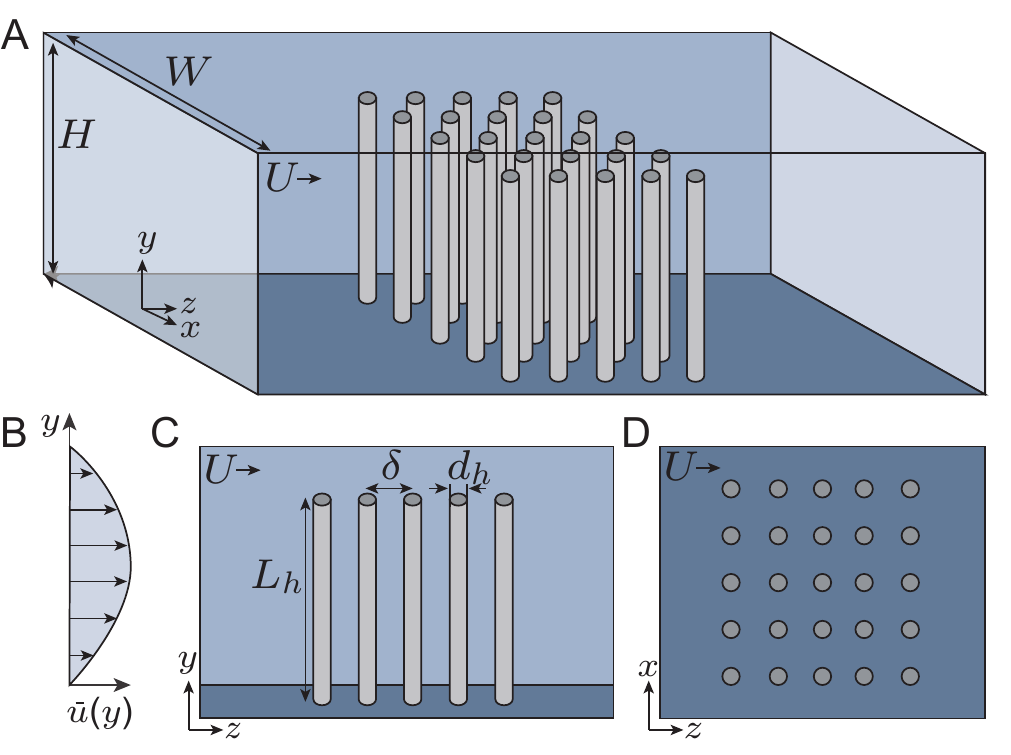}
	\caption{ (A) Channel diagram. (B) The undistrubed channel flow is a rectangular Poisuille flow $\ubar$. (C) Hairs have diameter $d_h$, length $L_h$, and center-to-center spacing $\delta$. (D) The hairs are arranged in a $5 \times 5$ rectangular grid. }\label{fig:diagram}
\end{figure}

\section*{Discovery of deflection phase}\label{sec:exp_res}

To investigate the physical principles, we developed an experimental model system of rigid hairs immersed in water (Fig. \ref{fig:diagram}A-D) We mounted rigid steel rods on one wall of a rectangular channel and visualize the flow with tracer particles. Features of the flow are set by the smallest length scale in this system, which is the diameter of the hairs, $d_h = 1$mm. We define the characteristic velocity to be the maximum flow speed $U$ in an undisturbed channel. Then, for water with density $\rho$ and viscosity $\mu$, we define the Reynolds number to be: $\Rey = \rho U d_h/\mu$.
We measure the flow phase as a function of $\Rey$ and the separation lengths $\delta$ of the hair bed by measuring the magnitude of the flow velocity in the center of the bed. The channel had cross-section dimensions $40$mm by $62$mm, and the separation lengths $\delta$ varied from $2$mm to $10$mm.

\begin{figure}[t]
	\centering
	\includegraphics[scale = 1]{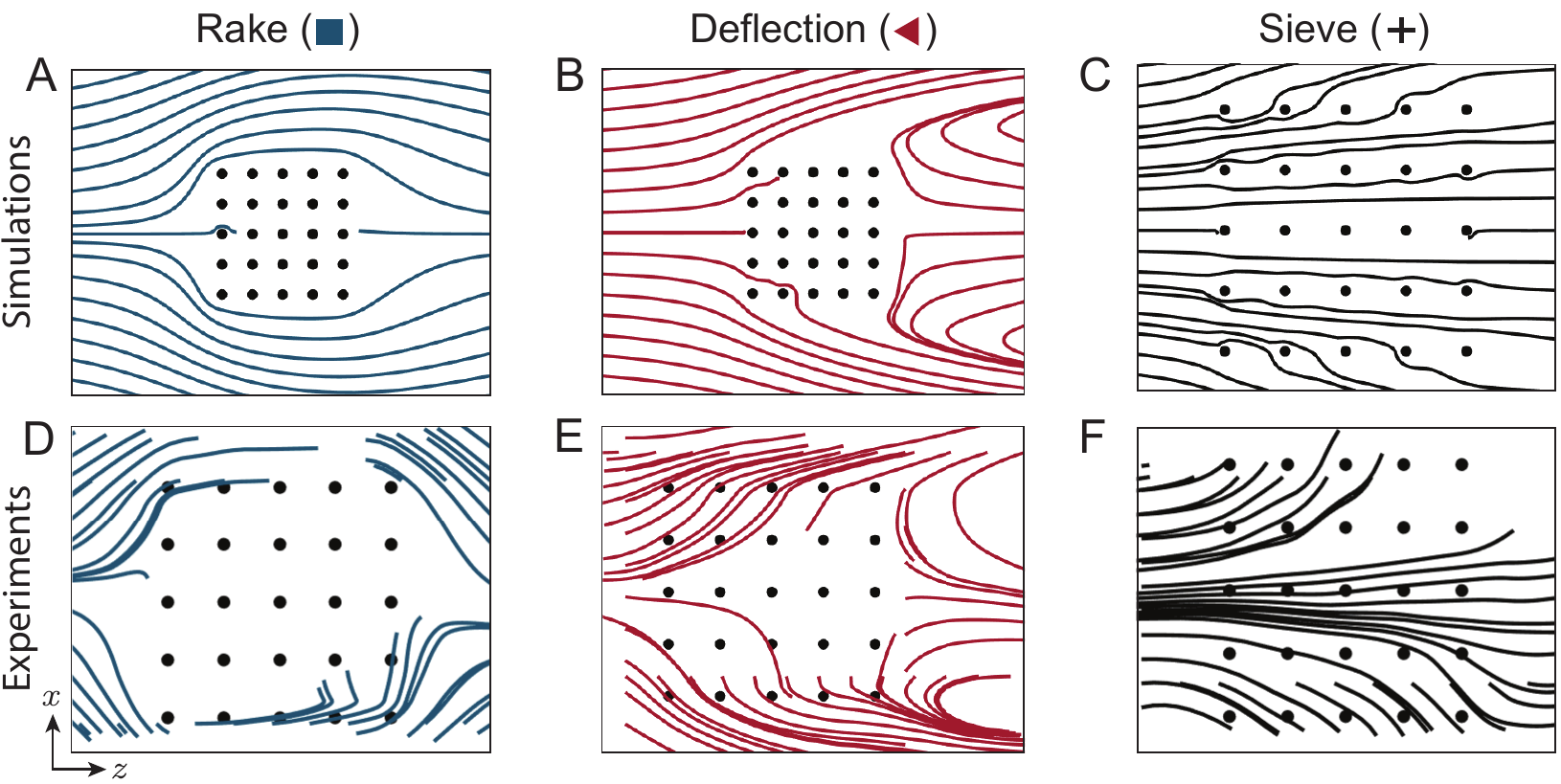}
	\caption{There are three phases of flow observed in experiment. (A) In the rake phase, numerical simulations show that streamlines circumvent the hair bed. (B) In the deflection phase, numerical simulations show that streamlines penetrate the hair bed but exit laterally. (C) In the sieve phase, numerical simulations show that the streamlines fully penetrate the hair bed. (D) Streamlines reconstructed from PIV measurements of the velocity show a rake at $\Rey = 0.8$ and $\delta=2$mm, black markers represent the centers of the cylinders.  (E) Streamlines reconstructed from PIV measurements of the velocity show the deflection at $\Rey = 31$ and $\delta = 2$mm.  (F) Streamlines reconstructed from PIV measurements of the velocity show the sieve at $\Rey = 19$ and $\delta = 4$mm.}\label{fig:phases}
\end{figure}

In the rake phase, streamlines circumvent the hair bed (Fig. \ref{fig:phases} A). The velocity magnitude inside the bed is an order of magnitude less than the undisturbed channel flow. The streamlines appear to have time reversal symmetry -- a characteristic of creeping flows ($\Rey \ll 1$) where, if time and velocity are reversed, the flow will evolve along the same streamlines. The rake phase can be observed at $\Rey = 0.8$ for the hair bed with separation length $\delta = 2$mm (Fig. \ref{fig:phases}D). This is consistent with the observations and simulations for crustaceans \cite{koehl2001fluid,cheer1987paddles}.

The sieve phase is characterized by streamlines that fully penetrate the hair bed (Fig. \ref{fig:phases} C). The streamlines move predominantly in the $z-$direction, with a slight deflection in the $x-$direction, thereby breaking time-reversal symmetry. The magnitude of velocities inside the bed are similar to that of the undisturbed channel flow. The sieve phase can be observed at $\Rey = 19$ for the hair bed with separation length $\delta = 4$mm (Fig. \ref{fig:phases}F). This is consistent with the observations and simulations for crustaceans \cite{koehl2001fluid,cheer1987paddles}.

While two phases of flow have been observed in crustaceans -- rake and sieve -- in our geometry we observe a third transitional region of flow which we call the \textit{deflection} phase. Here, fluid penetrates into the hair bed, but is deflected laterally out of bed. 

The deflection phase is characterized by streamlines with significant transverse displacement in the $x-$direction (Fig. \ref{fig:phases} B). The streamlines penetrate the hair bed, but are deflected transversely, exiting the bed before reaching the last row of hairs. We observe a circulation region in the wake of the hair bed. The velocity magnitude in the wake region is an order smaller than the undisturbed channel flow. Time reversal symmetry is clearly broken. The deflection phase can be observed at $\Rey = 31$ for the hair bed with separation length $\delta = 2$mm (Fig. \ref{fig:phases}E).

In order to design devices that can exploit these phases of flow, it would be useful to have a predictive theory that does not require a full numerical simulation. Given $\Rey$ and a hair bed separation length $\delta$, can we predict which phase the flow will exhibit? We conjecture that, to first order, the most significant predictor of the flow phase is the depth of the boundary layer on a single hair. In the next section, we develop a quantitative theory for boundary layer depth on a cylinder in a rectangular channel.

\section*{Boundary layer thickness around a single hair}\label{sec:1H}

A long slender body (like a hair or bristle) disturbs the surrounding flow due to the no-slip boundary condition on its surface. This gives rise to a region with a steep velocity gradient, called the boundary layer. The shape of the boundary layer depends on the geometry of the body and the Reynolds number of the flow: low-$\Rey$ flows have thick boundary layers and high-$\Rey$ flows have thin boundary layers.

\begin{figure}[t]
	\centering
	\includegraphics[scale = 0.88] {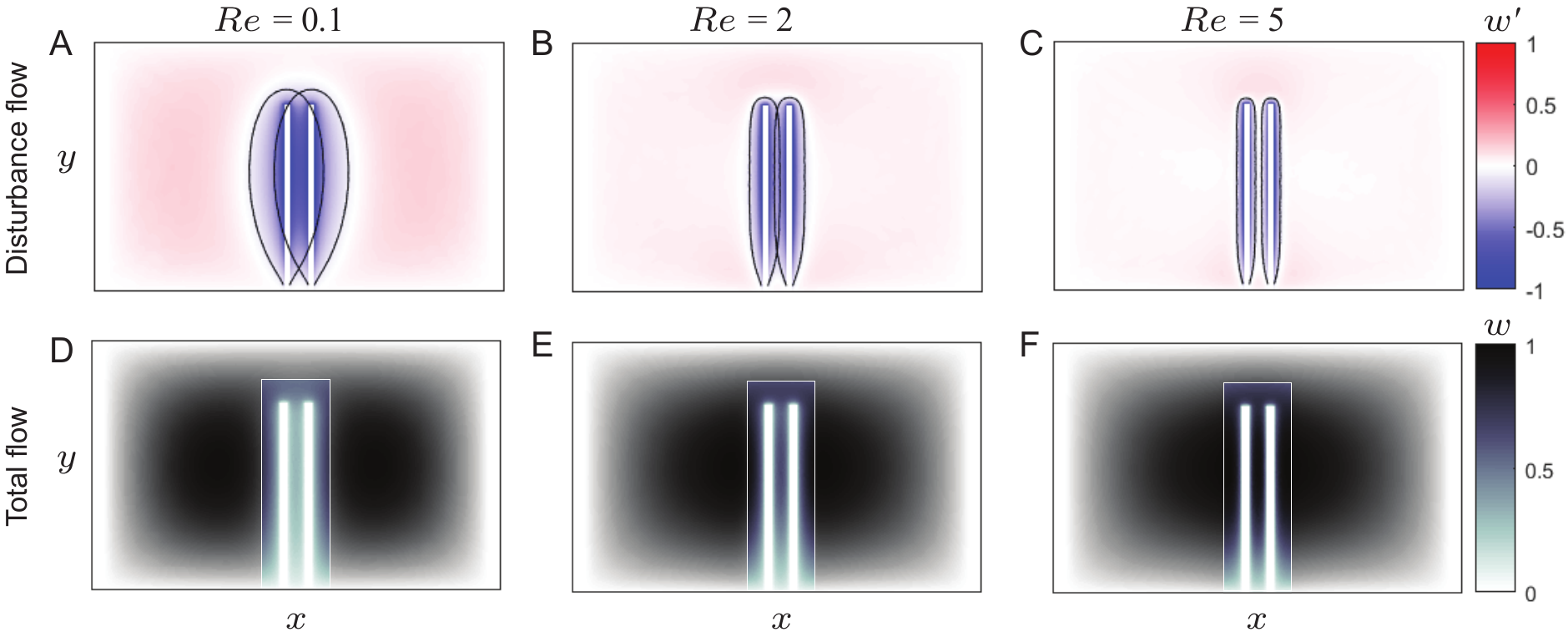}
	\caption{ The degree of overlap of the boundary layers determines the resulting flow phase. For all plots, the separation length is $\delta = 4mm$. The top row shows the disturbance flow $w'$ around two hairs, with the curve $\Gamma_p$ from a single hair super-imposed over each hair. The bottom row shows the absolute flow $w$ around the two hairs. At $\Rey = 0.1$, (A) the boundary layers overlap completely and (D) the total flow acts like a rake with stagnant flow in the gap between the hairs. At $\Rey = 2$, (B) the boundary layers partially overlap and (E) the total flow acts like a deflected flow with moderate speeds in the gap. And at $\Rey = 5$, (C) the boundary layers do not overlap, and (F) the total flow acts like a sieve, with large speeds in the gap.}\label{fig:gamma_contour_rc}
\end{figure}

For an array of hairs, we expect the flow to exhibit rake behavior when the boundary layers are so thick that they overlap in the region between the hairs (Fig. \ref{fig:gamma_contour_rc}A). Conversely, we expect the flow to exhibit sieve behavior when the boundary layers are too short to overlap (Fig. \ref{fig:gamma_contour_rc}E).

Therefore, in order to understand the flow around an array of rigid hairs, we first characterize the flow around a single hair. While the presence of neighboring hairs will affect the shape of the boundary layer non-linearly, these effects are higher order (since at $\Rey = O(1)$ the system is only weakly non-linear).   

\begin{figure}[t]
	\centering
	\includegraphics[scale = 1] {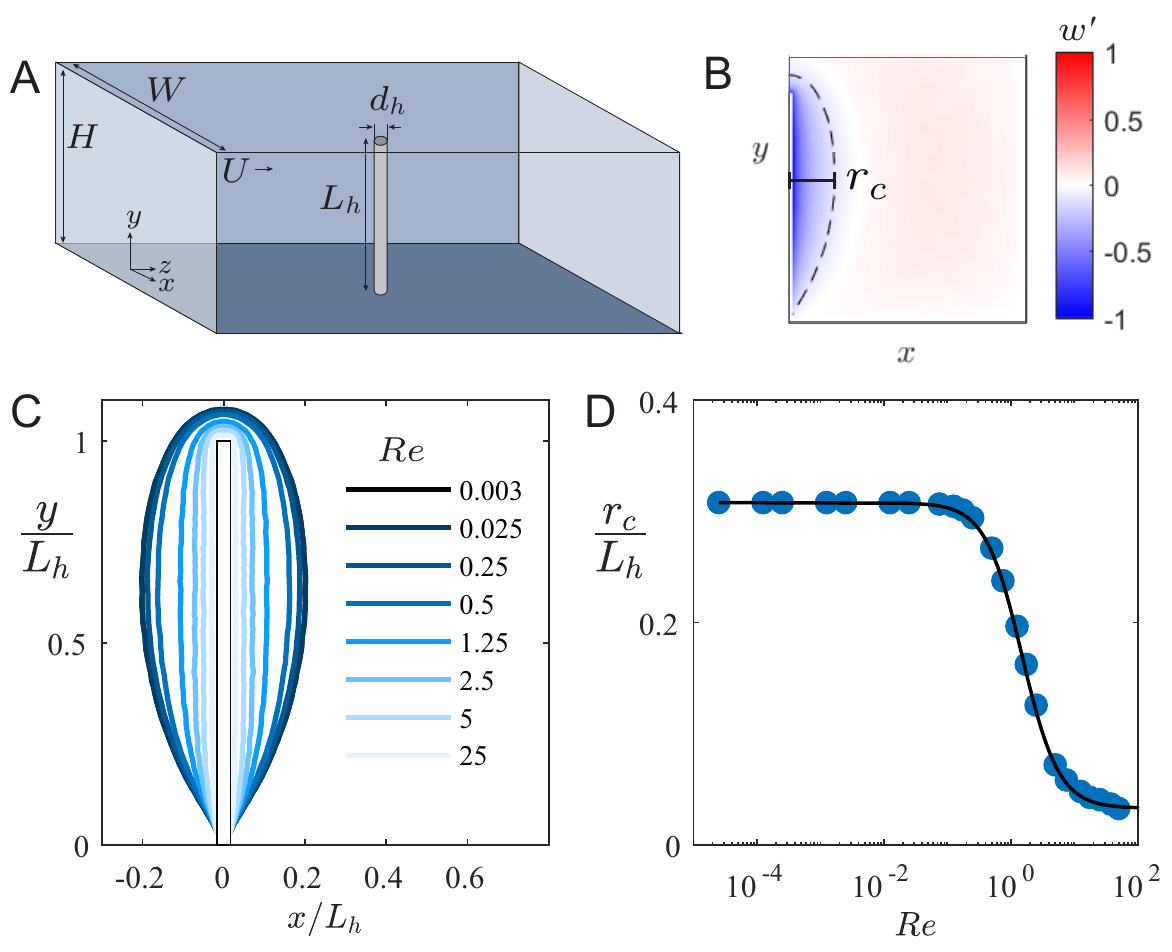}
	\caption{ (A) Diagram of a single hair in a rectangular channel. (B) The boundary layer is the negative (blue) region of the disturbance flow $w'$ at $\Rey = 0.025$.  (C) Plots of $\Gamma_p$ where $p = -0.1$ for various values of $\Rey$, here the black rectangle represents the hair. The critical radius $r_c$ as defined by equation (\ref{eq:rc}). (D)  The critical radius $r_c$ as a function of $\Rey$ is fit well by a logistic function (\ref{eq:rc_pois}) and shown as a solid black line. }\label{fig:gamma_rc}
\end{figure}

Consider a single cylinder anchored to one wall in a rectangular channel (Fig. \ref{fig:gamma_rc}A). The flow $\bu$ around the hair is described by the steady-state Navier-Stokes equations (\ref{eq:NSE1})--(\ref{eq:NSE2}) (Full derivation in SI). We assume that the flow at the inlet is the Poiseuille flow $\ubar$ for a rectangular channel \cite{Papanastasiou99}. It is useful to consider the normalized disturbance flow: $\bu' = (\bu -\ubar)/U$. 
Recall that $U$ is the characteristic velocity, which we determine to be the maximum velocity in an undisturbed channel. The resulting disturbance flow $\bu'$ has dimensionless units and therefore can be compared across $\Rey$.

The disturbance flow will take values between $-1 \le \bu' \le 0 + \epsilon$, where $\epsilon$ is a relative flow speedup. We observe speedup in this system due to the presence of the channel walls and conservation of mass. There is a large negative region of flow near the hair, which we call the boundary layer. In numerical simulations, we observe that at $\Rey = 0.025$ the boundary layer is thick and elliptical in shape (Fig. \ref{fig:gamma_rc}B). Whereas at a higher Reynolds number $\Rey = 2.5$, the boundary layer is thin and rectangular (Fig. \ref{fig:gamma_rc}C).

Further, we restrict our evaluation to the plane $z=0$ and only consider the streamwise component of the disturbance flow (i.e. for $\bu' = (u',v',w')$, consider only $w'$). Then we can measure the level-set with  value $p$ of the streamwise disturbance flow. We define the curve $\Gamma_p$ as follows:
\begin{equation}\label{eq:gamma}
\Gamma_p = \{ (x,y)\, \vert \,\, w'(x,y,0) = p \}\,.
\end{equation}

To determine the depth of the boundary layer, we consider the $p=-0.1$ level set. We define the critical radius $r_c$ as the x-component of the curve $\Gamma_p$ evaluated at the midway point of the channel $y = H/2$ (Fig. \ref{fig:gamma_rc}C).
\begin{equation}\label{eq:rc}
r_c = \Gamma_p \vert_{y = H/2} \, , \qquad p = -0.1\, .
\end{equation}


The critical radius is a decreasing function of $\Rey$ (Fig. \ref{fig:gamma_rc}D), and is well approximated by a logistic function in $\log_{10}{\Rey}$, specifically:
\begin{equation}\label{eq:rc_pois}
	\frac{r_c}{L_h} = m + \frac{M}{\left(\frac{\log_{10}\Rey}{\log_{10}\Rey_*}\right)^{k}+1}
\end{equation}
Equation (\ref{eq:rc_pois}) has four fitting parameters: $m = 0.026$, $M = 0.175$, $\Rey_* = 1.2$, and $k=1.5$.  

We can use this numerical fit of the critical radius to design hair beds to achieve either a rake, deflection, or sieve flow.

\section*{Designing arrays of rigid hairs}

We arrive at a simple design rule for predicting the phase of flow around a bed of rigid hairs. If the separation length $\delta < r_c$, then the boundary layers overlap in the gap between the hairs, which will result in a rake flow. Conversely, if $\delta > 2r_c$, then the boundary layers will not touch in the gap, which will result in a sieve flow. And  intermediate $\delta$ will result in a deflection phase.

\begin{figure}[t]
	\centering
	\includegraphics[scale = 1]{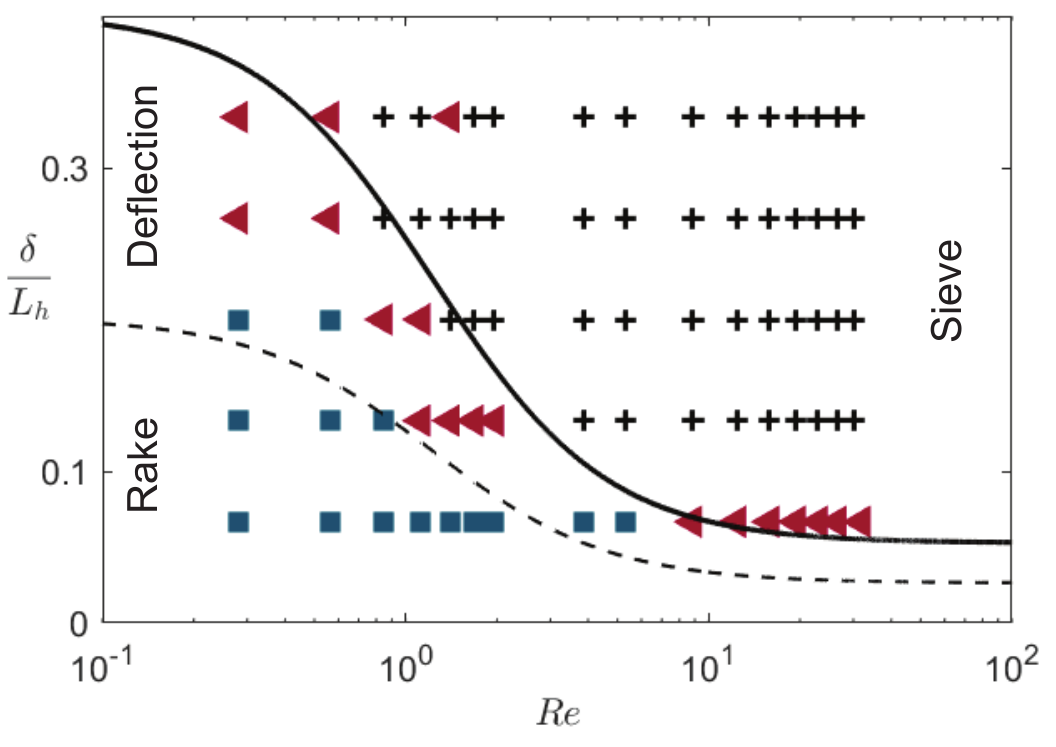}
	\caption{(A) Phase diagram of the observed flow: rake (blue square), deflection (red triangle), and sieve (black circle). The critical radius (\ref{eq:rc_pois}) determines the separation between rake flow and deflected flow (thin dashed line), while twice the critical radius determines the separation between deflected flow and sieve flow (thick solid line).}\label{fig:phase_data_rc}
\end{figure}

We can further verify this design rule by modeling the flow around two cylinders in a channel with a separation length $\delta = 4$mm. We observe that the curve $\Gamma$ predicting the boundary layer on a single cylinder in a channel is consistent with the boundary layers on two cylinders in a channel (Fig. \ref{fig:gamma_contour_rc}A,C,E). This shows that hydrodynamic interactions between hairs are higher order effects and validates our decision to ignore them at first order.

At $\Rey = 0.1$, the critical radius for a single hair is $r_c = 0.2L_h = 5.9$mm. Here, $r_c > \delta = 4$mm, so the boundary layer from each hair overlaps, covering the gap between the hairs (Fig. \ref{fig:gamma_contour_rc}A). According to our design rule, this configuration should produce a rake flow. We observe that the absolute flow is nearly stagnant in the gap region (Fig. \ref{fig:gamma_contour_rc}B), consistent with rake flow.

At $\Rey = 2$, the critical radius is $r_c = 0.08L_h = 2.5$mm. Here $r_c$ is comparable to the separation length $\delta = 4$mm. The boundary layer from each hair overlaps in the gap region, however the disturbance flow $w'$ is only moderately negative in the gap (Fig. \ref{fig:gamma_contour_rc}C). According to our design rule, this configuration should produce a deflection flow. We observe that the absolute flow has moderate speed in the gap region (Fig. \ref{fig:gamma_contour_rc}D), consistent with deflection flow.

At $\Rey = 5$, the critical radius $r_c = 0.05L_h = 1.4$mm is much smaller than the separation length $\delta$. The boundary layers are distinct with no overlap (Fig. \ref{fig:gamma_contour_rc}E). According to our design rule, this configuration should produce a sieve flow. We observe that the absolute flow is close to unity in the gap region (Fig. \ref{fig:gamma_contour_rc}F), consistent with sieve flow.

This rule is consistent with our experimental data (Fig. \ref{fig:phase_data_rc}A). Equation (\ref{eq:rc_pois}) for $r_c$ predicts the separation between the rake flow and the deflection flow, while the equation for $2r_c$ predicts the separation between deflection flow and sieve flow. The rake and sieve phases of flow agree qualitatively with 2D numerical predictions \cite{cheer1987paddles,chow1989drag}.


\section*{Hairy surfaces in crustaceans}

We compare our design space with that employed by the hairy appendages found in crustaceans. The hairy surfaces serve one of two purposes: chemo-sensing (stomatopod and lobster) and suspension-feeding (mole crab and barnacle). For each crustacean, the relevant measurements are the length of the hair $L_h$, the hair diameter $d_h$, the cross-stream spacing length $\delta_x$, the stream-wise spacing length $\delta_z$, and the $\Rey$ (Table \ref{tab:biodata}).

\begin{figure}[t]
	\centering
	\includegraphics[scale = 0.8] {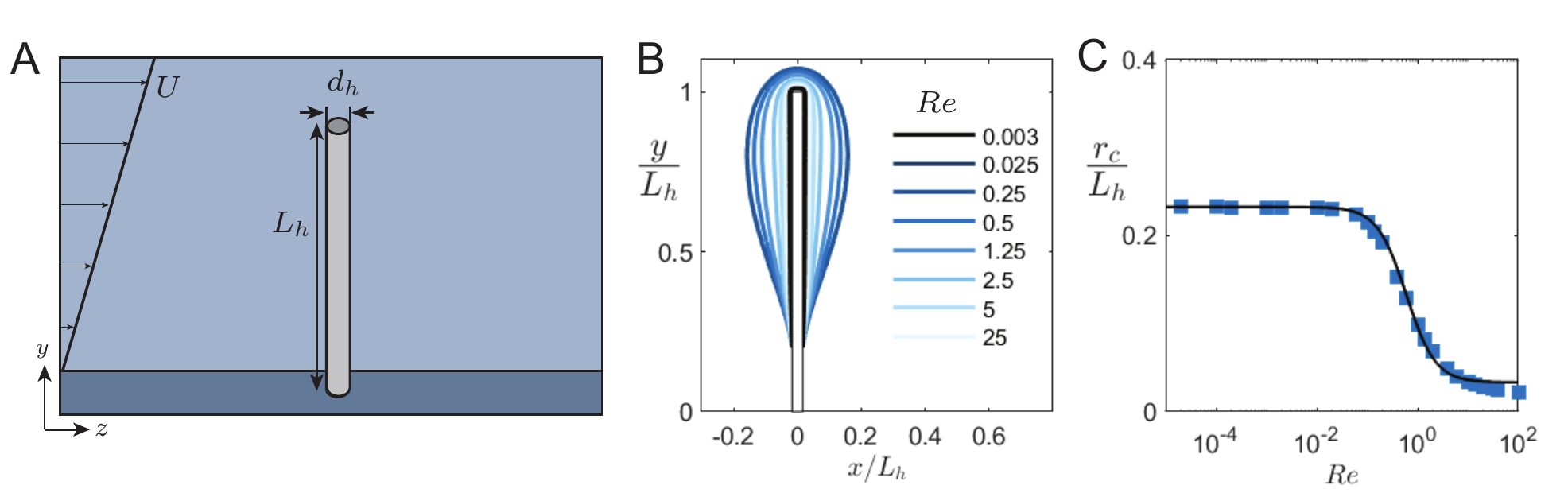}
	\caption{ (A) Diagram of a single hair in a rectangular channel subject to a shear flow.  (B) Plots of $\Gamma_p$ where $p = -0.1$ in a shear flow for various values of $\Rey$, here the black rectangle represents the hair. The critical radius $r_c$ as defined by equation (\ref{eq:rc}). (D)  In a shear flow, the critical radius $r_c$ as a function of $\Rey$ is fit well by a logistic function (\ref{eq:rc_pois}) and shown as a solid black line. }\label{fig:shear_gamma_rc}
\end{figure}

\begin{figure}[t]
	\centering
    \includegraphics[scale = 1]{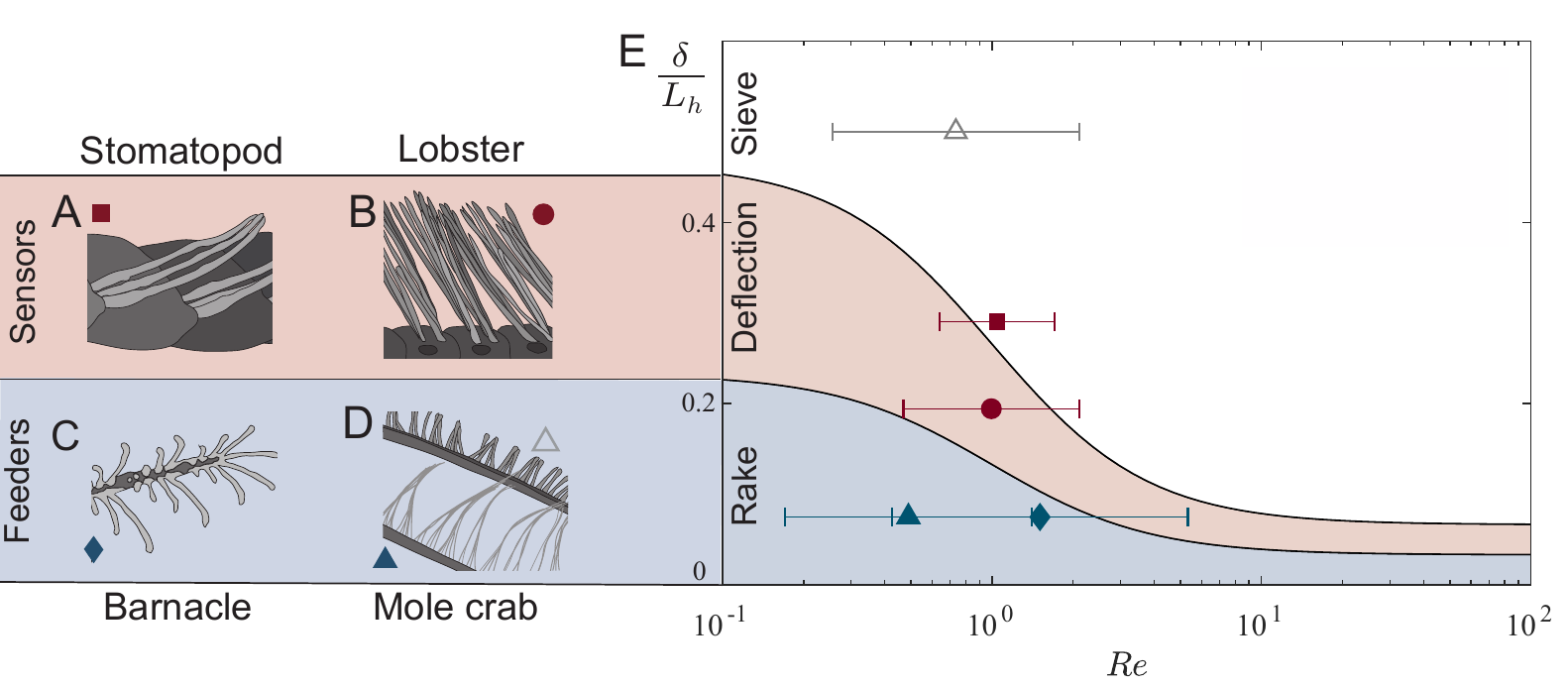}
	\caption{ (A) Stomatopod antennule, (B) lobster antennule, (C) barnacle cirra, and (D) mole crab setae. (E) The critical radius for a cylinder in an unbounded shear flow (\ref{eq:rc_pois}) determines the separation between rake flow and deflected flow (thin dashed line), while twice the critical radius determines the separation between deflected flow and sieve flow (thick solid line).  Boxes on the plot indicate ranges of $\Rey$ and $\delta/L_h$ observed in lobsters and stomatopods. }\label{fig:crustaceans}
\end{figure}

Marine crustaceans use their sense of smell to detect food, detect information about their neighbors, and avoid predators \cite{mead1999stomatopod}. Most crustaceans, including stomatopods and lobsters, detect odors using chemosensory sensillae called aesthetascs -- stiff hair-like structures located on the antennules. On the stomatopod \textit{Gonodactylaceus mutatus}, the aesthetascs are arranged in 16 rows, with each row consisting of three aesthetascs (Fig. \ref{fig:crustaceans}A). The aesthetascs form a $50^{\circ}$ angle with the surface of the antennule \cite{mead2000stomatopod}. Similarly, the spiny lobster \textit{Panulinus argus} has aesthetascs arranged in an array of 15 rows, with each row consisting of 10 aesthetascs \cite{reidenbach2008antennule}. The aesthetascs are angled laterally to form a $32^{\circ}$ angle with the antennule \cite{goldman2001fluid} and angled streamwise so that the tips form a zig-zag pattern (Fig. \ref{fig:crustaceans}B). Furthermore, the entire array of aesthetascs are surrounded by much larger guard hairs.

Suspension-feeders use a diverse range of filamentous appendages to capture particles from the surrounding water. For example, the barnacle \textit{Balanus glandula} feeding structure is called a cirral fan, consisting of paired biramous appendages covered with setae. Barnacles use their cirri to directly capture food, filter food from the water column, and reject non-nutritious items \cite{geierman2009feeding}. Barnacles are able to switch between active feeding (beating their cirri) and passive feeding (holding their cirri extended), in response to the current speed \cite{trager1990barnacle}. Each segment of the cirra has a two groups of six setae, arranged in order of decreasing length, and separated by an $47^{\circ}$ angle. For the purpose of this study, we consider the length of the second longest seta and the spacing length between the first and second seta (Fig. \ref{fig:crustaceans}C). The mole crab \textit{Emerita talpoida} uses a pair of second antenna to collect food particles, each antenna composed of long plumose flagella. Each flagellum is covered with duplicate sets of four rows of setae, which in turn are covered in shorter setules \cite{conova1999role}. The exteriormost setae are covered in fine, cylindrical setules, while the second pair of setae are covered in shorter, flat setules (Fig. \ref{fig:crustaceans}D).

Each crustacean moves its hairy appendage through the water, and therefore only the anchoring wall effects (in contrast to channel walls in our experiment) are prevalent in the model. So, instead of measuring the boundary layer depth $r_c$ on a hair in Poiseuille flow, we model $r_c$ in a shear flow (Fig \ref{fig:shear_gamma_rc}A-C). The critical radius is still defined by Eq (\ref{eq:rc}), and can be fit by the logistic function in Eq (\ref{eq:rc_pois}). For shear flow, the fitting parameters are: $m = 0.058$, $M = 0.253$, $\Rey_* = 3$, and $k=1.5$.  Note that the values of $m$, $M$, and $\Rey_*$ are significantly different than those for a cylinder in a rectangular channel. However, the design principle remains the same. If $\delta < r_c$, we predict a rake flow, if $\delta > 2r_c$ we predict sieve flow, and we predict deflection flow in the transition region.

We observe that the hairy appendages for the chemo-sensors (stomatopods and lobsters) lie in the deflection and sieve region of the diagram (Fig. \ref{fig:crustaceans}E). Conversely, the hairy appendages for suspension-feeders (mole crabs and barnacles) lie on the rake region of the diagram. One exception is that the short setules on the mole crab lie in the sieve region of the graph. However, this may be explained by the fact that the aspect ratio of the short setules $\kappa = 1/10$ is significantly different from the aspect ratio of the other crustacean hairs and the model $\kappa = 1/30$.

\begin{table*}[t]
\begin{center}
\begin{tabular}{ lrrrrrr } 
 Crustacean & $L_h$ & $d_h$ & $\kappa$ & $\delta_s$ & $\delta_{\ell}$ & $\Rey$\\ \hline
 Stomatopod (\textit{Gonodactylaceus mutatus} \cite{mead2000stomatopod}) & $516\mu$m & $20\mu$m & $1/26$ & $20\mu$m & $150\mu$m & 0.8 - 1.7 \\ 
 Lobster (\textit{Panulirus argus} \cite{goldman2001fluid}) & $720\mu$m & $22\mu$m & $1/33$ & $23\mu$m & $140 \mu$m & 0.5 - 2 \\ 
 Mole crab (\textit{Emerita talpoida} \cite{conova1999role}) & $200\mu$m & $4\mu$m & $1/50$ & - & $15\mu$m & 0.2 - 1.4 \\ 
Mole crab$^*$ (\textit{Emerita talpoida} \cite{conova1999role}) & $60\mu$m & $6\mu$m & $1/10$ & - & $30\mu$m & 0.3 - 2.1 \\ 
Barnacle (\textit{Balanus glandula} \cite{vo2018fluid}) & $363\mu$m & $10\mu$m & $1/36$ & - & $27\mu$m & 0.4 - 5.3 \\
\hline
\end{tabular}
\caption{Biological data for the dimensions of hairs on crustaceans. $^*$Mole crabs have different setae with longer setules (no star) and shorter setules (star). $^\dagger$Barnacles have different morphology depending on whether the surrounding water is stagnant (no dagger) or active (dagger).}\label{tab:biodata}
\end{center}
\end{table*}

\section*{Discussion}\label{sec:conc}

To summarize, we have shown that rectangular arrays of rigid hairs immersed in fluid flow exhibit separate phases of flow at different $\Rey$. Consistent with crustacean olfaction, we observe both the rake and sieve flow.  However, we uncovered a new intermediary phase, which we call the \textit{deflection} phase, where fluid partially penetrates the bed and escapes perpendicularly to the direction of flow.

The deflection phase has not been discussed in the literature to the best of our knowledge. This phase may not be observed in crustaceans because they have hair beds with different lateral and streamwise spacing lengths. For example, the stomatopod has a streamwise spacing length that is much smaller than the lateral spacing length (Table \ref{tab:biodata}). This may prevent streamlines from escaping laterally from the hair bed. Alternatively, it also could be a desirable feature for sensing. Since stomatopod hairs have chemo-receptors, it might be desirable to have fluid guide particles directly to the hairs.

Here we have shown that the phase of flow can be predicted form the depth of the boundary layer on a single hair.  We discovered a simple design principle based on the degree of overlap of the boundary layers in the gap between the hairs.  The theory for the depth of the boundary layers in equation (\ref{eq:rc_pois}) is specific to the geometry in this experiment. However, we have shown that the design principle is also valid for shear flow, and correlates with various marine crustacean hairy appendages, despite their varying geometries.

In particular, the flow phases described here depend on the fact that the hairs are tall relative to the channel, suppressing the effect of fluid escaping over the top of the hairs. Further work is needed to examine this behavior. Additionally, we have only considered hairs anchored normal to the surface of the channel. In lobsters the chemo-sensory hairs are angled so that the tips of the hairs form a zig-zag formation. Other angles of the hairs could affect the onset of the phases of flow, perhaps even directing the deflection flow to one side of the channel. Finally, we have not considered any deformation of the hairs. Deformation and reconfiguration is ubiquitous in biology at low-$\Rey$ flows, and can be observed at intermediate-$\Rey$ in the sensory appendages of the crab \textit{Callinectes sapidus}  \cite{gleeson1982morphological,reidenbach2008antennule}.

\section*{Materials and Methods}

Flow was observed in a 450 mm long channel fabricated from 1/4 inch acrylic sheets and joined with plastic cement (Scigrip). The channel cross-section dimensions were 62 mm $\times$ 40 mm ($W \times H$), respectively, with the shortest dimension identified as the depth ($y$) dimension (Fig. \ref{fig:diagram}A) and the longer dimension as the width dimension ($x$). The channel length was chosen to minimize the entry length effects in the hair bed (See SI). Fluid entered and exited the channel through 1/4 inch plastic barbed tube fittings (McMaster-Carr) placed 400mm apart (Fig. \ref{fig:diagram_experiment}). Six channels were constructed, one for each hair bed and one for empty channel flow.

The hair beds consisted of a $5 \times 5$ rectangular grid of cylinders perpendicular to the bottom surface of the channel (Fig. \ref{fig:diagram}A-D). Each cylinder was cut from 1 mm diameter steel rods (uxcell) to have an aspect ratio of $\kappa = 1/30$ comparable to aesthetacs observed in lobsters and stomatopods (Table \ref{tab:biodata}). The hair length $L_h = 30$mm was chosen to be close to the height $H$ in order to suppress the effect of fluid escaping over the hair bed. We constructed five hair beds and varied the center-to-center distance between the cylinders $\delta$ to be 2mm, 4mm, 6mm, 8mm, and 10mm. The cylinders were inserted into a laser-cut holes in an acrylic sheet and then secured to the bottom of each respective channel. The cylinders were coated with black spray paint (Rustoleum, flat black) to reduce light reflection.

The channel Reynolds number is defined as $\Rey_C = UH/\nu$, where $\nu = 1 \times 10^{-6} \mathrm{m}^2 \mathrm{s}^{-1}$ is the kinematic viscosity of deionized water at room temperature, $H = 40$mm is the short dimension of the channel, and $U$ is the maximum fluid velocity in the channel. The cylinder Reynolds number is defined at $\Rey = Ud_h/\nu$. Fifteen different total flow rates were used, including $Q$ = 20, 30, 40, 50, 60, 70 mL min$^{-1}$ and $Q$ = 2, 3, 5, 7, 9, 11, 13, 15, and 17.5 gph, corresponding to a range of channel Reynolds numbers $\Rey_C = 22.9 - 1260$ and cylinder Reynolds numbers $\Rey = 0.56 - 31 $.

Fluorescent red polyethylene microspheres (Cospheric) were used to visualize the flow. The particles had 25 $\mu$m diameter and density of $0.995$g/cc. The particles were dispersed at 0.004 volume fraction in a suspending fluid composed of deionized water and 0.001 (wt/vol) biocompatible surfactant (Tween 80). The particles were pumped into the channel in one of two ways. At low speeds, the particles were pumped at a controlled flow rate using a multichannel syringe pump (New Era NE-1600). At high speeds, the particles were pumped using a submersible 80gph water pump (Songjoy). The flow rate was measured at the output with an acrylic flowmeter (McMaster-Carr). The solutions were infused using 1/4 inch plastic tubing (McMaster-Carr).

The particles were chosen to be near-neutrally buoyant with a particle density of $0.995$g cm$^{-3}$. The particle density does not match the density of the suspending fluid (density 1.000g cm$^{-3}$).  The sedimentation velocity can be determined by balancing buoyancy force with the drag force for a sphere. For this experiment, the sedimentation velocity is $1.6\mu$m/s, meaning that the particles sediment a distance between $0.02 - 2$ mm, depending on the $\Rey$. 
At high speeds, the sedimentation can be ignored in this experiment, but sedimentation can be observed at lower speeds. 

The fluorescent particles were illuminated by two 532nm 50mW green lasers fixed with a cylindrical lens attachment to flatten the laser line into a sheet (Laserland). The laser sheets were arranged perpendicularly to illuminate the plane $y=0.5H$ (Fig. \ref{fig:diagram}A-D). A support structure made of extruded aluminum was constructed to fix the location of the lasers. 

Particle velocities were tracked by imaging using a camera (Phantom Miro M320S). At low speeds we used 24 fps with 40ms exposure time, and at high speeds we used 100 frames per second with 10ms exposure time.  The camera was focused to the plane illuminated by the laser sheet.
The effective pixel size ranged between $14\mu$m and $20\mu$m in each video.

\subsection*{Determining particle velocities}

Videography provided measurements of the $z$ and $x$ (stream-wise and lateral) velocities of particles located in the plane $y = 0.5H$ illuminated by the laser sheet. Since the particles particle fluoresce red light, we filter the color videos to retain only the red channel data. Then we use the particle image velocimetry (PIV) code MatPIV \cite{matpiv} to develop a vector field representing the displacements of all particles from one frame to the next. 

\begin{figure}[t]
	\centering
    \includegraphics[scale = 1] {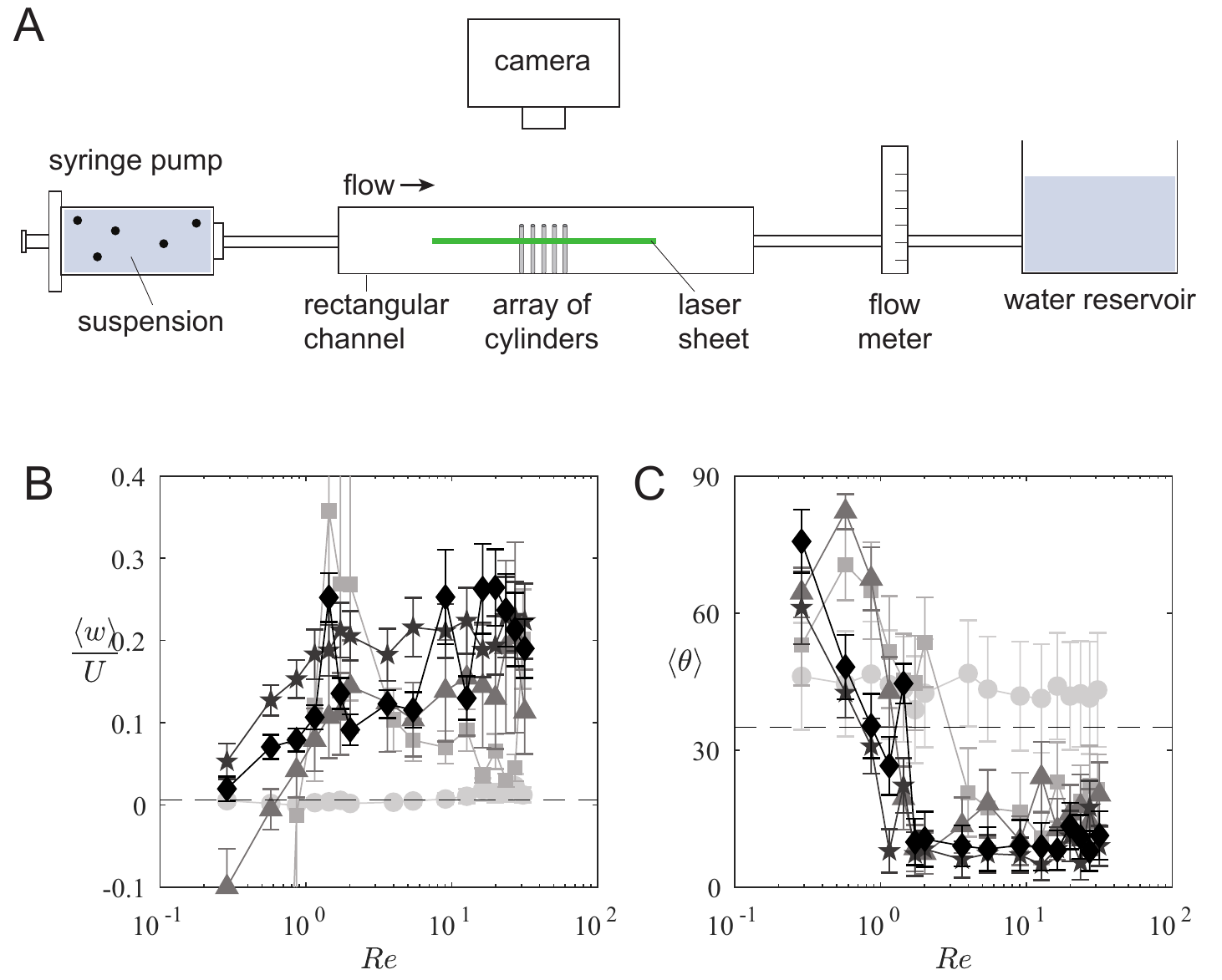}
	\caption{(A) The experimental setup. (B) Measurement of the normalized average streamwise velocity $\langle w \rangle / U$ can be used to determine whether or not the flow is a rake. Error bars represent standard error, while circles mark $\delta = 2$mm, squares mark $\delta = 4$mm, triangles mark $\delta = 6$mm, stars mark $\delta = 8$mm, and diamonds mark $\delta = 10$mm. (C) measurement of the average angle $\langle \theta \rangle$ distinguished between deflection and sieve flow.} \label{fig:diagram_experiment}
\end{figure} 

Because the laser-sheet is located on one side of the channel, the intensity of the fluorescent particles declines as they move away from the light source. Additionally, there are significant shadows generated by the cylinders. In order to compensate, we measure the velocities of particles over five frames in order to sample enough particles moving through the shaded region of the channel.

To classify the flow phase, we restrict the domain to the space between two rows of hairs (i.e. $0 < x < \delta$ and $-2.5\delta < z <2.5\delta$. We compute the average stream-wise and transverse velocities ($w$ and $u$ respectively) over all of the particles in the starting frame that are located in $S_i$. To account for the sedimentation of particles in the $x$-direction, we subtracted off the sedimentation velocity $v_s = 1.7\mu$m/s off of the velocity $u$. The flow phase can be quantified by the average downstream velocity $\langle w \rangle$ inside the the hair bed (Fig. \ref{fig:diagram_experiment}B). 
When $\langle w \rangle < 0.006 U$, then the flow inside the hair bed is essentially stagnant, representing the rake flow phase. Then at larger velocities, we determine whether the flow is a deflection or a sieve by calculating the average angle $\langle \theta \rangle$ (Fig. \ref{fig:diagram_experiment}C). At angles less than $35^{\circ}$ the flow is mostly stream-wise, which we categorize as sieve flow, and at larger angles the flow is deflecting laterally in the deflection phase. 

\subsection*{Numerical Simulation}
We model the flow around a long cylinder attached at one end to a rectangular channel (Fig. \ref{fig:gamma_rc}A). Let the origin lie in the center of the channel. A cylinder with height $L_h = 30$mm and diameter $d_h = 1$mm is attached to the bottom of the channel at location $(x,y,z) = (0, -\frac{1}{2}H, 0)$.  The long axis of the cylinder extends in the $y-$direction. The aspect ratio of the hair is $\kappa = d_h/L_h = 1/30$, similar to measurements of crustaceans (Tab. \ref{tab:biodata}). 
Let $\bu$ be the flow around the hair and $p$ the corresponding pressure. Then the equations of motion are the steady state Navier-Stokes equations:
\begin{subequations}
\begin{align}
 \mu \nabla^{2} \bu - \nabla p &= \rho \, \bu \cdot \nabla \bu \,, \label{eq:NSE1} \\
 \nabla \cdot \bu & = 0 \,, \label{eq:NSE2} 
 \end{align}
\end{subequations}
Subject to the boundary conditions:
 \begin{subequations}
\begin{align}
\bu & = 0 \mbox{ on cylinder }  \sqrt{x^{2} + z^{2}} = \frac{d_h}{2}, \quad y \le L_h - \frac{H}{2}\,, \label{eq:NSE3} \\
 \bu &= 0 \mbox{ on channel walls } x = \pm \frac{W}{2}, y = \pm \frac{H}{2} \,, \\
 \bu &= \ubar \mbox{ as } z \to \pm \infty \,. \label{eq:NSE5}
\end{align}
\end{subequations}
We solve the equations (\ref{eq:NSE1})--(\ref{eq:NSE2}) and (\ref{eq:NSE3})--(\ref{eq:NSE5}) numerically using COMSOL Multiphysics (Cambridge, MA) with 188,159 elements.  Accuracy of this model is evaluated in the SI.

Additionally, we model the flow around a cylinder attached to a wall in a shear flow. We define the shear to be $\bu_{\gamma} = (0,0,u_{\gamma})$, where: $u_{\gamma} = \gamma (y+.5H)$.
We define the Reynolds number as: $\Rey = \rho \gamma L_h d_h/\nu$.
Then the flow $\bu$ and corresponding pressure $p$ solve the steady-state Navier-Stokes equations (\ref{eq:NSE1})--(\ref{eq:NSE2}) with the following boundary conditions:
\begin{subequations}
\begin{align}
 \bu &= 0 \mbox{ on bottom wall } y = - \frac{H}{2} \,, \label{eq:NSE3g} \\
 \bu &= \gamma H \mbox{ on top wall } y = + \frac{H}{2}\,, \\
 \bu &= \bu_{\gamma} \mbox{ as } z \to \pm \infty \,, \\
 [-p \mathbf{I} + \mu(\nabla \bu &+ (\nabla \bu)^T]\cdot \mathbf{n} = 0 \mbox{ on } x = \pm \frac{W}{2} \,.\label{eq:NSE5g}
\end{align}
\end{subequations}
We solve these equations (\ref{eq:NSE1})--(\ref{eq:NSE2}) and (\ref{eq:NSE3g})--(\ref{eq:NSE5g}) numerically using COMSOL Multiphysics (Cambridge, MA) with 158,262 elements.

\section*{Acknowledgements}
K.H. was supported by National Science Foundation Grant DMS-1606487. We'd like to thank Anoop Rajappan for assistence in the fabrication of PIV setup.


\providecommand{\noopsort}[1]{}\providecommand{\singleletter}[1]{#1}%
%

\widetext
\begin{center}
\textbf{\large Supplemental Materials: Marine crustaceans with hairy appendages: role of hydrodynamic boundary layers in sensing and feeding}\\
Kaitlyn Hood, M. S. Suryateja Jammalamadaka, A. E. Hosoi
\end{center}
\setcounter{equation}{0}
\setcounter{figure}{0}
\setcounter{table}{0}
\setcounter{page}{1}
\makeatletter
\renewcommand{\theequation}{S\arabic{equation}}
\renewcommand{\thefigure}{S\arabic{figure}}
\renewcommand{\thetable}{S\arabic{table}}
\renewcommand{\thesection}{S\arabic{section}}
\renewcommand{\bibnumfmt}[1]{[S#1]}
\renewcommand{\citenumfont}[1]{S#1}

\section{Experimental considerations}\label{sec:exp_des}

For each experiment we define two Reynolds numbers: the channel Reynolds number $\Rey_C$ and the hair Reynolds number $\Rey$. If $\rho$ and $\nu$ are the density and viscosity of water, respectively, $H = 40$mm is the height of the channel, $d_h = 1$mm the diameter of the hair, and $U$ the maximum Poisuielle flow in the channel, then 
\begin{equation}
\Rey_C = \frac{\rho UH}{\nu} \,, \qquad \Rey = \frac{\rho Ud_h}{\nu} \,.
\end{equation}

To calculate the Reynolds numbers of the channel, we measured the maximum velocity $U_m$ in an empty channel at each flow rate. In Table \ref{tab:sedimentation} we list the flow rates $Q$, measured maximum velocity $U_m$, and the Reynolds numbers $\Rey$ and $\Rey_C$.

\subsection{Sedimentation Calculation}

\begin{table}[b]
\begin{center}
\begin{tabular}{llll}
$Q$ (mL/min) & $\Rey$ & $\Rey_C$ & $d_s$ (mm) \\ \hline
20           & 0.6    & 23       & 0.59       \\
30           & 0.8    & 34       & 0.393      \\
40           & 1.1    & 46       & 0.295      \\
50           & 1.4    & 57       & 0.236      \\
60           & 1.7    & 69       & 0.197      \\
70           & 2      & 80       & 0.169      \\
139          & 3.9    & 159      & 0.085      \\
189          & 5.3    & 216      & 0.062      \\
315          & 8.8    & 361      & 0.037      \\
442          & 12.3   & 505      & 0.027      \\
568          & 15.8   & 649      & 0.021      \\
694          & 19.4   & 793      & 0.017      \\
820          & 22.9   & 937      & 0.014      \\
946          & 26.4   & 1082     & 0.012      \\
1104         & 30.8   & 1262     & 0.011     \\ \hline
\end{tabular}
\caption{Flow rate, Reynolds numbers, and sedimentation distance of tracer particles for each $\Rey$.}\label{tab:sedimentation}
\end{center}
\end{table}

We calculate the sedimentation velocity of the tracer particles by balancing buoyancy force with the viscous drag on a sphere. Let $a$ be the radius of the particles, $\Delta \rho$ be the difference in density between the water and the particle, $g$ the gravity constant, then this force balance is:
\begin{equation}
\frac{4}{3} \pi \Delta\rho g a^3 = 6 \pi \mu a v_s \,.
\end{equation}
Rearranging for the sedimentation velocity yields:
\begin{equation}
v_s = \frac{2 \Delta\rho g a^2}{9 \mu} = 1.7\mu \mathrm{m/s} \,.
\end{equation}

In experiments, the maximum velocity in the channel ranged from $U = 0.6 - 70$mm/s, corresponding to Reynolds numbers $\Rey = 0.6 - 31$. Since the channel was $L = 400$mm long, at the center of the channel (and the center of the hair bed) we expect particles to sediment a distance:
\begin{equation}
d_s = \frac{v_s L}{2U}  = 0.6-0.01 \mathrm{mm}. 
\end{equation}
A full table of the sedimentation distance for each experiment is listed in Table \ref{tab:sedimentation}.


\subsection{Poiseuille development length} \label{app:entry_length}

Poiseuille flow in a rectangular channel has an entry length $L_e$ from the channel inlet where the flow is not fully formed. A general rule of thumb for the entry length $L_e$ for a rectangular microchannel is $L_e = \frac{1}{30} \Rey_C W$ \cite{ciftlik2013high}. For our channel, this becomes roughly $L_e \sim 2 \Rey_C$ mm = $46 - 2524$mm.

\begin{figure}[b]
	\centering
	\includegraphics[scale = 0.5]{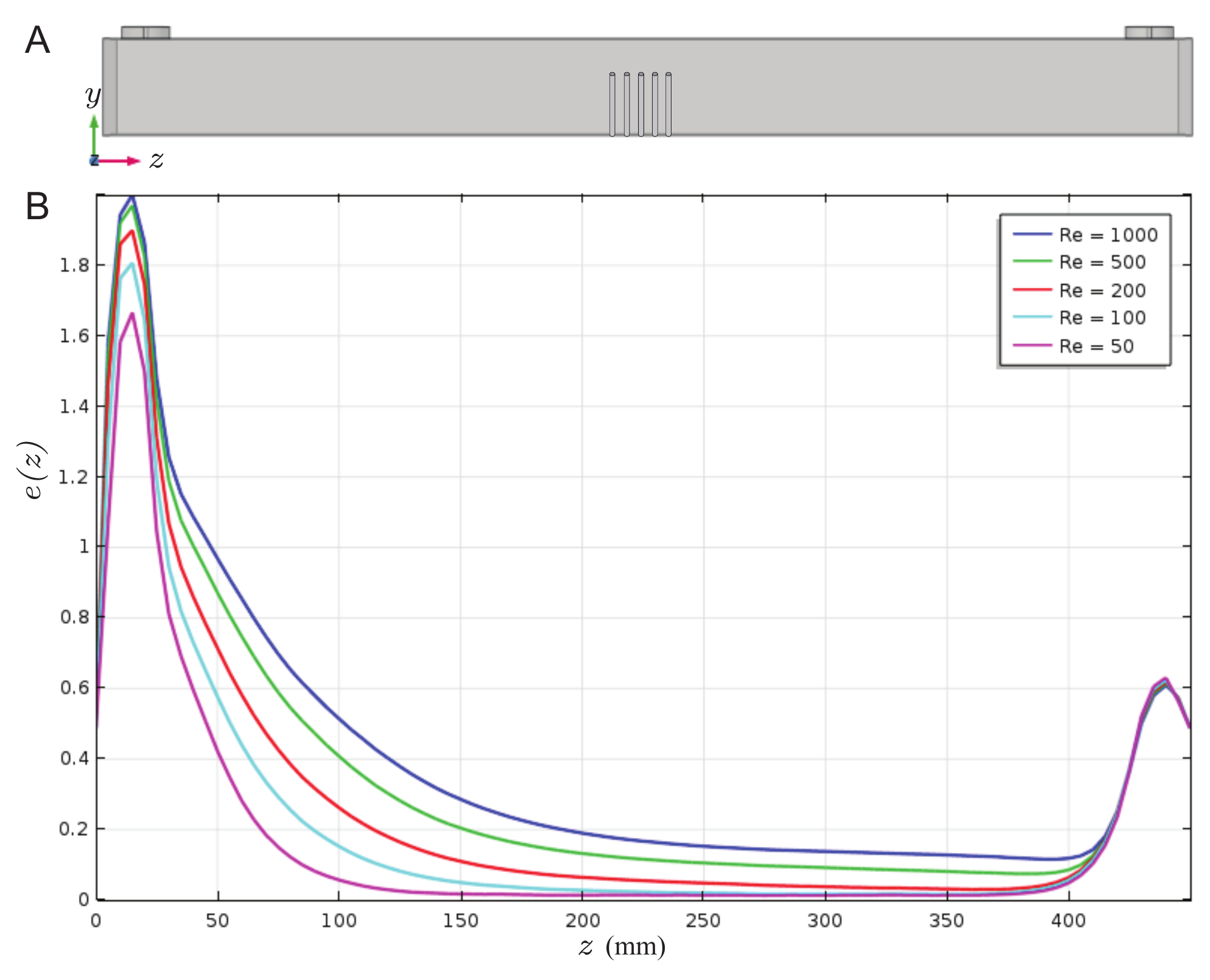}
	\caption{(A) Diagram of channel with location of hair bed illustrated. (B) Error $e(z)$ between empty channel flow and Poiseuille flow as a function of $\Rey_C$.  }\label{fig:error}
\end{figure}

To ensure that the channel has fully-developed Poiseuille flow in the hair bed, we simulate flow through a rectangular channel using FEM with 75,608 elements in Comsol Multiphysics (Cambridge, MA). The channel has dimensions $450$mm $\times 62$mm $\times 40$mm with flow entering and exiting through cylinders on the top of the channel placed $400$mm apart (Figure \ref{fig:error}A).

The background flow, $\ubar$, is rectangular channel Poiseuille flow \cite{ Papanastasiou99}, and takes the form $\ubar=  \bar{u}(x,y)\mathbf{e}_{z}$, where $\bar{u}$ is defined by:
\begin{align}\label{eq:ubar}
\begin{split}
  \bar{u}(x,y) &= A \Bigg[  -\frac{1}{2} \left(\left(\frac{y}{H}\right)^2 - \left(\frac{1}{2}\right)^2\right) 
  - \frac{4}{\pi^3} \sum_{n=0}^\infty  \frac{ (-1)^n \cosh\left(\lambda_n x\right)\cos\left(\mu_n y\right)}{(2n+1)^3  \cosh\left(\lambda_n W /2 \right)}  \Bigg]\,, \\ 
\lambda_n &= \frac{(2n+1)\pi}{W} \,, \qquad \mu_n = \frac{(2n+1)\pi}{H} \,, \qquad  \qquad A = 0.0736 U \,.
\end{split}
\end{align}
The velocity $\mathbf{\bar{u}}$ and pressure $\bar{p}$ solve the Stokes equations with boundary condition $\mathbf{\bar{u}} = \mathbf{0}$ on the channel walls.

We define the error between the simulated flow $\bu' = (u',v',w')$ and rectangular Poiseuille flow $\ubar = (0,0,\bar{u})$ as:
\begin{equation}
	e(z) = \frac{1}{UHW} \iint  \sqrt{ [u'(x,y,z)]^2 + [v'(x,y,z)]^2 + [w'(x,y,z) - \bar{u}(x,y)]^2 } \, \mathrm{d}x \, \mathrm{d}y \,.
\end{equation}
Recall that $U$ is the maximum flow velocity in the Poiseuille flow, and $U$ depends on $\Rey_C$. For all $\Rey_C$, the error $e$ is large at the inlet, decreases to near zero in the center of the channel, and increases again at the outlet (Figure \ref{fig:error}B). The hair bed is placed in the center of the channel, and the simulation predicts that the error will be less than $10\%$ for $\Rey_C \le 200$. Furthermore, the flow should have less than $20\%$ error in the hair bed for flows up to $\Rey_C =1000$.

\section{Experimental Data}

\begin{figure}
	\centering
	\includegraphics[scale=1]{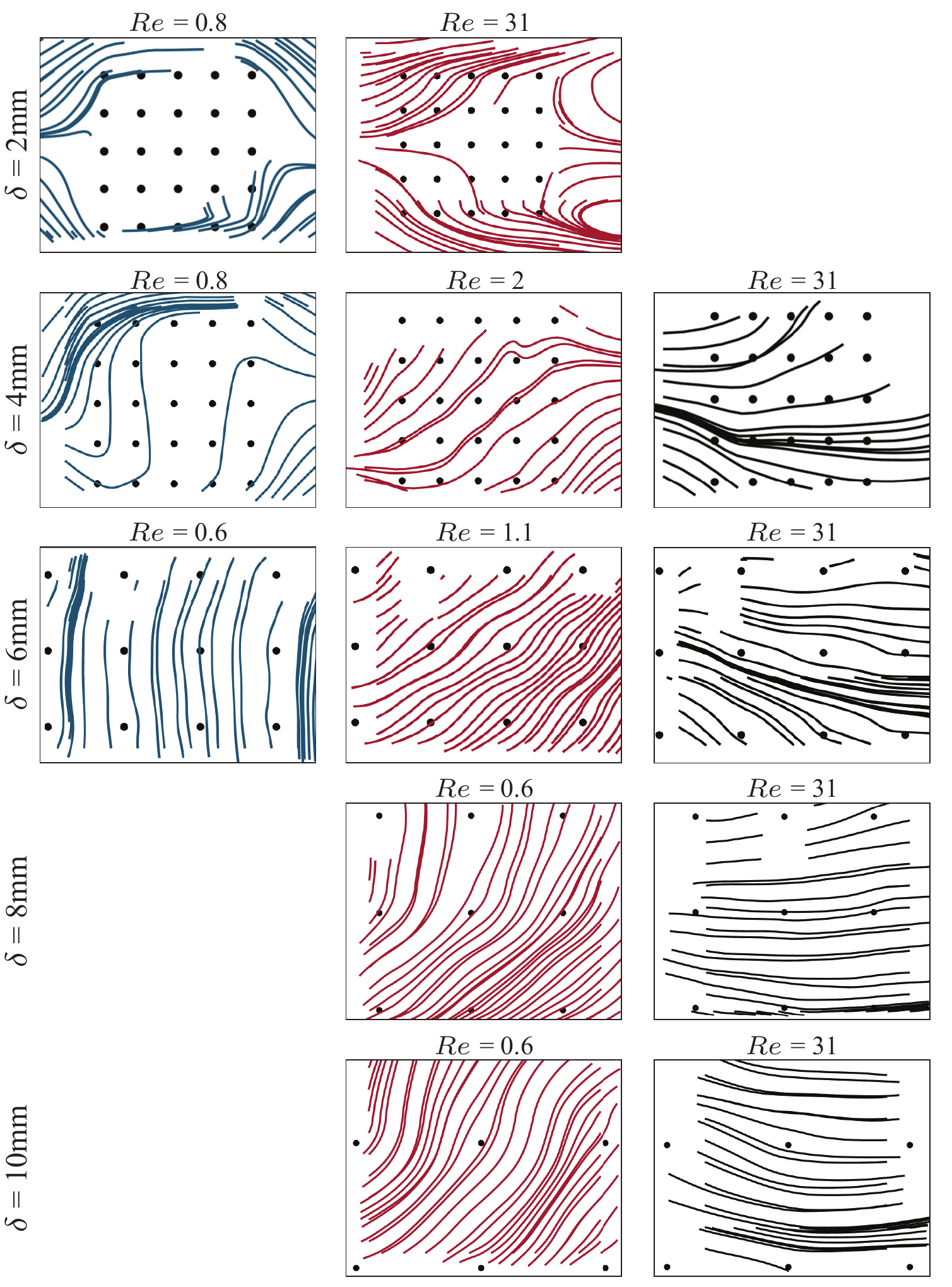}
	\caption{Streamlines from PIV analysis of video dat for each separation length $\delta$. Flows in the rake phase are shown in the left column, flows in the deflection phase are shown in the middle column, and flows in the sieve phase are shown in the right column.}\label{fig:piv1}
\end{figure}

We run particle-image velocimetry (MatPIV \cite{matpiv}) on each video from the experiment. Visually, we can qualitatively identify each flow as either rake, sieve, or deflection (Fig. \ref{fig:piv1}).

However, we can also quantitatively determine the flow phase by measuring the average stream-wise flow speed $\langle w \rangle$ and the average flow angle $\langle \theta \rangle$ inside the hair bed (Tab. \ref{tab:data1} - \ref{tab:data2}).

\begin{table}
\begin{center}
\begin{tabular}{lllll}
$\delta$ (mm) & $\Rey$ & $\langle w \rangle$    & $\langle \theta \rangle$ & phase \\ \hline
2             & 0.3    & 0.005  & 46       & 0     \\
2             & 0.6    & 0.002  & 45       & 0     \\
2             & 0.9    & 0      & 47       & 0     \\
2             & 1.1    & 0.003  & 44       & 0     \\
2             & 1.4    & 0.004  & 43       & 0     \\
2             & 1.7    & 0.006  & 39       & 0     \\
2             & 2      & 0.002  & 42       & 0     \\
2             & 4      & 0.003  & 47       & 0     \\
2             & 5.4    & 0.005  & 43       & 0     \\
2             & 9      & 0.007  & 42       & 1     \\
2             & 12.7   & 0.011  & 41       & 1     \\
2             & 16.3   & 0.019  & 44       & 1     \\
2             & 19.9   & 0.015  & 42       & 1     \\
2             & 23.5   & 0.015  & 42       & 1     \\
2             & 27.1   & 0.024  & 41       & 1     \\
2             & 30.8   & 0.013  & 43       & 1     \\ \hline
4             & 0.3    & -0.084 & 52       & 0     \\
4             & 0.6    & -0.09  & 70       & 0     \\
4             & 0.9    & -0.002 & 64       & 0     \\
4             & 1.1    & 0.025  & 51       & 1     \\
4             & 1.4    & 0.088  & 45       & 1     \\
4             & 1.7    & 0.083  & 44       & 1     \\
4             & 2      & 0.11   & 53       & 1     \\
4             & 4      & 0.066  & 21       & 2     \\
4             & 5.4    & 0.081  & 17       & 2     \\
4             & 9      & 0.12   & 16       & 2     \\
4             & 12.7   & 0.233  & 11       & 2     \\
4             & 16.3   & 0.184  & 23       & 2     \\
4             & 19.9   & 0.334  & 15       & 2     \\
4             & 23.5   & 0.153  & 19       & 2     \\
4             & 27.1   & 0.231  & 22       & 2     \\
4             & 30.8   & 0.207  & 21       & 2     \\ \hline
6             & 0.3    & -0.102 & 64       & 0     \\
6             & 0.6    & -0.006 & 82       & 0     \\
6             & 0.9    & 0.043  & 68       & 1     \\
6             & 1.1    & 0.081  & 43       & 1     \\
6             & 1.4    & 0.11   & 19       & 2     \\
6             & 1.7    & 0.114  & 9        & 2     \\
6             & 2      & 0.078  & 7        & 2     \\
6             & 3.6    & 0.113  & 13       & 2     \\
6             & 5.4    & 0.106  & 18       & 2     \\
6             & 9      & 0.142  & 10       & 2     \\
6             & 12.7   & 0.159  & 24       & 2     \\
6             & 16.3   & 0.148  & 13       & 2     \\
6             & 19.9   & 0.133  & 16       & 2     \\
6             & 23.5   & 0.197  & 18       & 2     \\
6             & 27.1   & 0.209  & 14       & 2     \\
6             & 31.7   & 0.116  & 20       & 2     \\ \hline
\end{tabular}
\caption{Data measurement values from PIV.}\label{tab:data1}
\end{center}
\end{table}

\begin{table}
\begin{center}
\begin{tabular}{lllll}
$\delta$ (mm) & $\Rey$ & $\langle w \rangle$    & $\langle \theta \rangle$ & phase \\ \hline
8             & 0.3    & 0.054  & 61       & 1     \\
8             & 0.6    & 0.13   & 43       & 1     \\
8             & 0.9    & 0.157  & 31       & 2     \\
8             & 1.1    & 0.188  & 8        & 2     \\
8             & 1.4    & 0.193  & 22       & 2     \\
8             & 1.7    & 0.218  & 7        & 2     \\
8             & 2      & 0.211  & 8        & 2     \\
8             & 3.6    & 0.187  & 6        & 2     \\
8             & 5.4    & 0.221  & 7        & 2     \\
8             & 9      & 0.217  & 7        & 2     \\
8             & 12.7   & 0.23   & 5        & 2     \\
8             & 16.3   & 0.194  & 8        & 2     \\
8             & 19.9   & 0.199  & 11       & 2     \\
8             & 23.5   & 0.234  & 5        & 2     \\
8             & 27.1   & 0.235  & 17       & 2     \\
8             & 31.7   & 0.229  & 9        & 2     \\ \hline
10            & 0.3    & 0.039  & 76       & 1     \\
10            & 0.6    & 0.157  & 49       & 1     \\
10            & 0.9    & 0.162  & 36       & 2     \\
10            & 1.1    & 0.219  & 27       & 2     \\
10            & 1.4    & 0.259  & 45       & 1     \\
10            & 1.7    & 0.278  & 10       & 2     \\
10            & 2      & 0.281  & 11       & 2     \\
10            & 3.6    & 0.252  & 9        & 2     \\
10            & 5.4    & 0.296  & 8        & 2     \\
10            & 9      & 0.259  & 9        & 2     \\
10            & 12.7   & 0.266  & 9        & 2     \\
10            & 16.3   & 0.269  & 8        & 2     \\
10            & 19.9   & 0.271  & 13       & 2     \\
10            & 23.5   & 0.243  & 11       & 2     \\
10            & 27.1   & 0.219  & 8        & 2     \\
10            & 31.7   & 0.195  & 11       & 2     \\ \hline
\end{tabular}
\caption{Data measurement values from PIV (cont'd).}\label{tab:data2}
\end{center}
\end{table}

\section{Equations of motion}

We model the flow around a long cylinder attached at one end to a rectangular channel.  The rectangular channel has dimensions $W=62$mm wide ($x$-axis) by $H=40$mm tall ($y$-axis) by $L = 400$mm long ($z-$axis). Let the origin lie in the center of the channel 

Inspired by the stomatopod and lobster biology (Table 
, we model the hair as a cylinder with aspect ratio $\kappa = 1/30$. A cylinder with height $L_h = 30$mm and diameter $d_h = 1$mm is attached to the bottom of the channel at location $(x^*,y^*,z^*) = (0, -\frac{1}{2}H, 0)$.  The long axis of the cylinder extends in the $y-$direction. We define the aspect ratio of the hair to be $\kappa = d_h/L_h$. The objective of this paper is to calculate the size of the boundary layer on the cylinder.

Let $\bu$ be the flow around the hair and $p$ the corresponding pressure. Then the equations of motion are the steady state Navier-Stokes equations:
\begin{subequations}
\begin{align}
 \mu \nabla^{2} \bu - \nabla p &= \rho \, \bu \cdot \nabla \bu \,, \label{eq:NSE1} \\
 \nabla \cdot \bu & = 0 \,, \\
 \bu & = 0 \mbox{ on cylinder }  \sqrt{x^{2} + z^{2}} = \frac{d_h}{2}, \quad y \le L_h - \frac{H}{2}\,, \\
 \bu &= 0 \mbox{ on channel walls } x = \pm \frac{W}{2}, y = \pm \frac{H}{2} \,, \\
 \bu &= \ubar \mbox{ as } z \to \pm \infty \,. \label{eq:NSE5}
\end{align}
\end{subequations}
In comsol, the outlet boundary condition is represented as a zero pressure flux BC.

Solve these equations (\ref{eq:NSE1})--(\ref{eq:NSE5}) numerically using COMSOL Multiphysics (Cambridge, MA) with 188,159 elements.  Accuracy of this model is evaluated in Appendix \ref{app:accuracy}.

\subsection{Shear flow}

For comparing boundary layer depth to crustacean measurements, we simulate a cylinder in a shear flow. We define the shear to be $\bu_{\gamma} = (0,0,u_{\gamma})$, where:
\begin{equation}
u_{\gamma} = \gamma (y+.5H) \,.
\end{equation}
And here, we define the Reynolds number as
\begin{equation}
\Rey = \frac{\rho \gamma L_h d_h}{\nu}\, .
\end{equation}
That is to say, the shear gradient starts at zero at $y = -.5H$ and reaches a velocity of $U_{\gamma} = \gamma L_h$ at the tip of the hair. It's this velocity $U_{\gamma}$ that we use as the characteristic velocity.

Let $\bu$ be the flow around the hair and $p$ the corresponding pressure. Then the equations of motion are the steady state Navier-Stokes equations:
\begin{subequations}
\begin{align}
 \mu \nabla^{2} \bu - \nabla p &= \rho \, \bu \cdot \nabla \bu \,, \label{eq:NSE1g} \\
 \nabla \cdot \bu & = 0 \,, \\
 \bu & = 0 \mbox{ on cylinder }  \sqrt{x^{2} + z^{2}} = \frac{d_h}{2}, \quad y \le L_h - \frac{H}{2}\,, \\
 \bu &= 0 \mbox{ on bottom wall } y = - \frac{H}{2} \,, \\
 \bu &= \bu_{\gamma} \mbox{ as } z \to \pm \infty \,, \\
 \bu &= \mbox{no flux condition at top and side walls} \,.\label{eq:NSE5g}
\end{align}
\end{subequations}
In comsol, the outlet boundary condition is represented as a zero pressure flux BC.

Solve these equations (\ref{eq:NSE1g})--(\ref{eq:NSE5g}) numerically using COMSOL Multiphysics (Cambridge, MA) with 158,262 elements.  Accuracy of this model is evaluated in Appendix \ref{app:accuracy}.

\section{Accuracy of numerical model}\label{app:accuracy}

\begin{figure}[b]
	\centering
	\includegraphics[scale = 1]{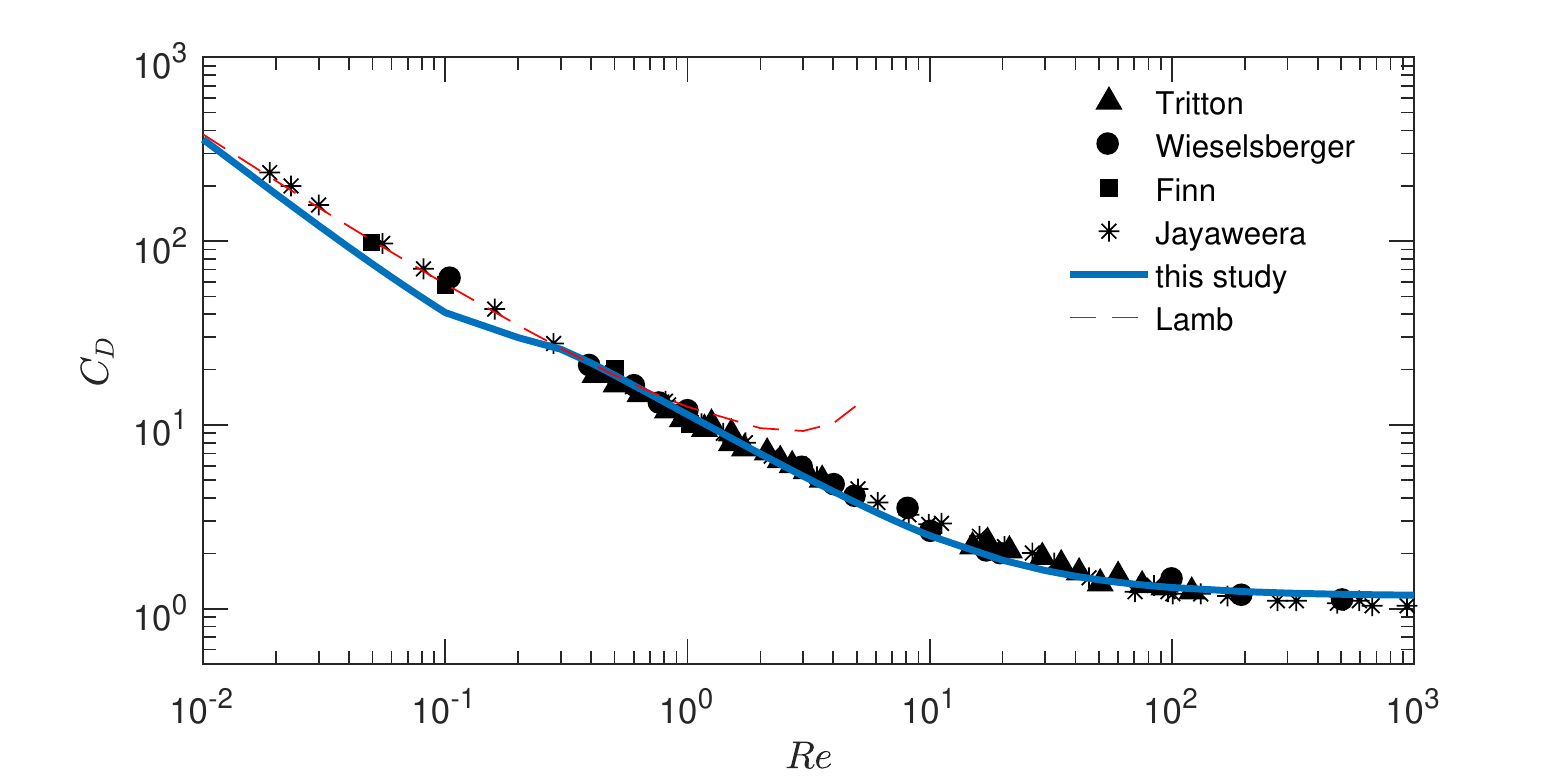}
	\caption{The drag coefficient on a cylinder as a function of Reynolds number. Experimental data is represented by markers ($\blacktriangle$ - Tritton \cite{tritton1959experiments}, $\bullet$ - Wieselsberger \cite{wieselsberger1922further}, $\blacksquare$ - Finn \cite{finn1953determination}, and $*$ - Jayaweera and Mason \cite{jayaweera1965behaviour}), the drag computed in this study by a solid blue line, and Lamb's formula (\ref{eq:lamb}) by a dashed red line. }\label{fig:dragvRe}
\end{figure}

To test the accuracy of our numerical method we compute the drag on a falling sphere and compare it to experimental data.  We use Comsol Multiphysics (Cambridge, MA) with 82,385 elements.  We consider a cylinder with length $L$ and radius $a = \kappa L$, where the aspect ratio is $\kappa = 0.05$.  The cylinder is assumed to be oriented with the long axis parallel to the ground.  The cylinder is sedimenting in a box with dimensions $10L \times 10L \times 10L$.  The background flow has density $\rho$, kinematic viscosity $\nu$, and is moving with velocity $U$ while the cylinder is moving with velocity $U_p = \frac{1}{3}U$.  The boundary condition on the surface of the cylinder is chosen to satisfy the weak constraints in order to achieve higher accuracy.  Then the Reynolds number of this system is defined as $\Rey = U_p \kappa L/ \nu$.


We compute the drag force $F_D$ on the cylinder by integrating the weak constraint Lagrange multiplier over the surface of the cylinder \cite{hood15inertial}. Then the drag coefficient $C_D$ is defined by:
\begin{equation}
C_D = \frac{F_D}{\rho U_p^2 a L} = \frac{F_D}{\rho U_p^2 \kappa L^2}
\end{equation}
Additionally, we observe Lamb's formula for the drag on a cylinder at low Reynold's number \cite{Lamb45}, explicitly:
\begin{equation}\label{eq:lamb}
C_{\infty} = \frac{8\pi}{(\frac{1}{2} - \gamma - \ln \frac{1}{8} \Rey_{\infty})\Rey_{\infty}}
\end{equation}

The computed drag coefficient agrees well with experimental data \cite{jayaweera1965behaviour, tritton1959experiments, wieselsberger1922further, finn1953determination}, especially at Reynolds numbers larger than unity (Figure \ref{fig:dragvRe}).  Around $\Rey = 0.1$ there seems to be a small negative bump in the numerical computation.  Otherwise, the numerical computation in this study agrees well with the data and with Lamb's formula (\ref{eq:lamb}).



\providecommand{\noopsort}[1]{}\providecommand{\singleletter}[1]{#1}%

\end{document}